\documentclass{elsart}

\usepackage{amssymb}
\usepackage{amsmath}
\usepackage{bm}
\usepackage{latexsym}
\usepackage{graphicx}
\usepackage{times}
\DeclareMathAlphabet{\mathpzc}{OT1}{pzc}{m}{it}
\DeclareMathAlphabet{\mathcalligra}{T1}{calligra}{m}{n}

\begin{document}

\begin{frontmatter}
\title{An existence of universality \\in the dynamics of two types of glass-forming liquids \\- fragile liquids and strong liquids}

\author{Michio Tokuyama}
\address{Institute of Multidisciplinary Research for Advanced Materials, Tohoku University, Sendai 980-8577, Japan}

\date{\today}

\begin{abstract}
By employing a simplified nonlinear memory function proposed recently by the present author, a universal equation for a collective-intermediate scattering function derived based on the time-convolutionless mode-coupling theory is numerically solved to study the dynamics of glass-forming liquids. The numerical calculation is based on the simulation results performed on two types of liquids, fragile liquids and strong liquids. The numerical solutions are then shown to be uniquely determined by the long-time collective diffusion coefficient $D(q_m)$, where $q_m$ is a first peak position of a static structure factor for a whole system. It is confirmed based on four different simulation results that the supercooled state consists of two substates, a weakly supercooled state in which the nonlinear parameter $\mu$ in the memory function increases as $D$ decreases and a deeply supercooled state in which $\mu$ becomes constant up to the glass transition. Here $\mu$ is shown to be constant in a liquid state, while it is shown to grow rapidly in a glass state. The value of $\mu$ in each state is shown to be uniquely determined by $D$ and also to depend on a type of liquids. Hence there exists such a universality that there is only one solution for different liquids of a same type at a given value of $D$. This may be consistent with the fact that strong liquids are structurally quite different from fragile liquids. Thus, it is emphasized that such a universality must be helpful to predict $q_m$ from experimental data.
\end{abstract}

\begin{keyword}
Glass transition; Simplified nonlinear memory function; Supercooled liquids; Time-convolutionless mode-coupling theory; Universality

\maketitle
\end{keyword}
\end{frontmatter}

\section{Introduction}

In order to investigate the dynamics of glass-forming materials near the glass transition from a unified statistical-mechanical point of view, we have recently proposed the time-convolutionless mode-coupling theory (TMCT) from first principles \cite{toku14} and thus derived the ideal TMCT equation for the collective-intermediate scattering function \cite{toku15}. The nonlinear memory function contained in this equation has exactly the same form as that obtained by the mode-coupling theory (MCT) \cite{mct84B,mct91}, which is written in terms of the static structure factor. Hence this equation can be numerically solved if such a static structure factor is known. In fact, it has been solved \cite{kim16,toku19} by using the Percus-Yevick static structure factor \cite{py}. Thus, it has been shown that the critical volume fraction $\phi_c\simeq 0.5817$ agrees with that predicted from the molecular-dynamics simulations performed on hard spheres \cite{toku03,toku031,toku05,toku071} and also that the numerical solutions describe the simulation results well within error, except at the $\beta$-relaxation stage \cite{toku19}. However, the static structure factor is in general not known and can be found numerically either from simulation results nor from experimental data. In the previous paper \cite{toku20}, therefore, we have reasonably simplified the nonlinear memory function and then solved the TMCT equation numerically based on the simulation results performed on two types of  glass-forming liquids, fragile liquids and strong liquids. We have taken the binary mixtures A$_{80}$B$_{20}$ with the Stillinger-Weber potential (SW) \cite{sw} as a typical example of fragile liquids and SiO$_2$ with the Nakano-Vashishta potential (NV) \cite{nv} as a simple example of strong liquids. Then, we have suggested that there exists only one solution in each type of liquids if a scaled collective diffusion coefficient $D$ has the same value in different liquids of a same type. In the present paper, we check whether such a universality holds for other liquids, such as Al$_2$O$_3$ with the Born-Meyer potential \cite{bm} and SiO$_2$ with the Beest-Kramer-Santen (BKS) potential \cite{bks} or not. Thus, we confirm that there exists each universality in each type of liquids. Hence we emphasize that such a universality must be used to predict a first peak position $q_m$ of the static structure factor for a whole system. This may be helpful to understand the structure of liquids near the glass transition experimentally. In fact, it is not easy to find $q_m$ experimentally, although it is easy to obtain numerically by simulations.

\section{Basic equations}
We consider the three-dimensional equilibrium glass-forming system, which consists of $N$ particles with mass $m$ and diameter $\sigma$ in the total volume $V$ at temperature $T$. Let $\xi$ denote the control parameter, such as a volume fraction $\phi(=\pi\rho \sigma^3/6)$ and an inverse temperature $1/T$, where $\rho(=N/V)$ is the number density. As shown in \cite{toku15}, the ideal TMCT equation for the cumulant function $K(q,t)(=-\ln[f(q,t])$ is given by
\begin{equation}
\frac{\partial^2 K(q,t)}{\partial t^2}=\frac{q^2v_{th}^2}{S(q)}-\zeta_0\frac{\partial K(q,t)}{\partial t}
-\int_0^t\Delta\varphi(q,t-s)\frac{\partial K(q,s)}{\partial s}ds, \label{ieq}
\end{equation}
where $f(q,t)$ is a scaled collective-intermediate scattering function with $f(q,0)=1$, $\zeta_0$ a friction coefficient, $S(q)$ a static structure factor, and $v_{th}$ an averaged thermal velocity. Here $\Delta\varphi(q,t)$ is a nonlinear memory function which has exactly the same form as that obtained by MCT \cite{mct84B,mct91}. The numerical solutions are expected to describe the simulation results well within error, except at the $\beta$-relaxation stage \cite{toku19}. As discussed in Ref. \cite{toku171}, such a disagreement at $\beta$ stage is just because a kind of Markov approximation has been used to derive Eq. (\ref{ieq}) from an original TMCT equation whose solutions are expected to describe a whole relaxation process. In the present paper, we use Eq. (\ref{ieq}) instead of the original one because the original one is quite difficult to solve even numerically \cite{toku171}. From Eq. (\ref{ieq}), one can easily find the asymptotic solutions as $K(q,t)\simeq q^2v_{th}^2 t^2/(2S(q))$ for a short time and $ K(q,t)\simeq q^2D_c(q)t$ for a long time, where $D_c(q)$ is a $q$-dependent collective diffusion coefficient
\begin{equation}
D_c(q)=\frac{v_{th}^2/S(q)}{\zeta_0+\int_0^{\infty}\Delta\varphi(q,s)ds}. \label{dcf} 
\end{equation}
The existence of the critical point $\xi=\xi_c$ in Eq. (\ref{ieq}) has been confirmed \cite{toku15,toku171}. By employing the same mathematical approach as that used in MCT \cite{mct91}, therefore, one can also write $D_c(q)$ in terms of a singular function as $D_c(q,\xi)\propto (1-\xi/\xi_c)^{\gamma_c}$, where $\gamma_c$ is a power exponent to be determined.

There exist two types of glass-forming liquids, (F) fragile liquids and (S) strong liquids \cite{ang91,soko93,boh93,vil93,ang95,ang00,deb01,ngai02,to}. This has been well-known for a long time since Angell \cite{ang88} has proposed a famous classification in viscosities of glass-forming materials. As shown in Refs. \cite{toku13,toku16}, the mean-$n$th displacement in a liquid coincides with the other mean-$n$th displacements in different liquids of a same type at a given value of a universal parameter $u_s(=-\log(D_sq_m/v_{th}))$, where $D_s$ is a long-time self-diffusion coefficient and $q_m$ a first peak position of $S(q)$. However, we note that the displacement in (F) never coincides with that in (S), even if $u_s$ has the same value. In general, these situations also hold for the cumulant function $K(t)(=K(q_m,t))$ \cite{toku20}. Hence we now derive a universal equation for $K(t)$ to describe the dynamics of glass-forming liquids, where the control parameter is an inverse temperature $1/T$. Since Eq. (\ref{ieq}) depends on the physical quantities $q_m$, $S(q_m)$, and $v_{th}$, it is convenient to introduce the time scale $\tau_D$ by $\tau_D=S(q_m)^{1/2}/(q_mv_{th})$. Using a scaled time $\tau=t/\tau_D$, one can then transform Eq. (\ref{ieq}) into a dimensionless equation \cite{toku20}
\begin{equation}
\frac{\partial^2 K(\tau)}{\partial \tau^2}=1-\zeta\frac{\partial K(\tau)}{\partial \tau}
-\kappa\int_0^t M(\tau-s)\frac{\partial K(s)}{\partial s}ds, \label{kasc}
\end{equation}
where $\zeta=\zeta_0\tau_D$ and $M(\tau)=\Delta\varphi(q_m,\tau)/\Delta\varphi(q_m,0)$. Here the coupling parameter $\kappa$ is given by $\kappa=\tau_D^2\Delta\varphi(q_m,0)$. This is a universal equation to discuss an existence of a universality in each type of liquids.
One can also obtain the asymptotic solutions as $K(\tau)\simeq\tau^2/2$ for $\tau\ll 1$ and $\tau/\tau_L$ for $\tau>\tau_{\alpha}$
with the diffusion time $\tau_L(=D^{-1})$, where $D(T)$ is the scaled collective diffusion coefficient given by
\begin{equation}
D(T)=q_m^2\tau_D D_c(q_m). \label{sltdc}
\end{equation}

By simply assuming that $S(k)$ obeys a Gaussian distribution with a peak position $k=q_m$, we have recently simplified the scaled memory function $M(\tau)$ reasonably as \cite{toku20}
\begin{equation}
M(\tau)=\frac{f(\tau)}{[1+\mu K(\tau)]^{\beta/b}}, \label{almod}
\end{equation}
where $\mu$ is a nonlinear parameter to be determined, $b$ the von Schweidler exponent \cite{vsc}, and $\beta$ the Kohlrausch-Williams-Watts (KWW) exponent \cite{kw,ww}, where the numerical values of $b$ and $\beta$ are listed in Table \ref{table-1} \cite{toku191}. Here we note that $\mu$ must play a role of a nonlinear exponent in $f(\tau)$ since one can write $\mu K(\tau)$ as $\mu K(\tau)=-\ln[f(\tau)^{\mu}]$.
\begin{table}[b]
\caption{Time exponents $b$ and $\beta$ for different type of liquids \cite{toku191}.}
\begin{center}
\begin{tabular}{cccc}
\hline
System  & $b$ & $\beta$ & $\beta/b$ \\
\hline
control parameter $1/T$ &&&\\
Fragile liquids &  0.3064 & 0.6832 & 2.230\\
Strong liquids &  0.2779 & 0.6809 & 2.451\\
\hline
\end{tabular}
\end{center}
\label{table-1}
\end{table}

\section{Numerical solutions}
In order to solve Eq. (\ref{kasc}) with the memory function given by Eq. (\ref{almod}) numerically, one needs to fix the values of three unknown parameters $\zeta$, $\kappa$, and $\mu$ consistently. As shown in the previous papers \cite{toku15,toku20toku17}, this is done by using the simulation results at each value of $D(T)$. In fact, the friction coefficient $\zeta_0$ has been found to be constant from the short-time behavior of the simulation results \cite{toku17}. Hence one can fix the value of $\zeta$ uniquely at each value of $D$. By using Eqs. (\ref{dcf}) and (\ref{sltdc}), one can write the coupling parameter $\kappa$ as
\begin{equation}
\kappa=\frac{1-\zeta D}{D \int_0^{\infty}M(\tau) d\tau}. \label{kappa}
\end{equation}
\begin{figure}
\begin{center}
\includegraphics[width=8.5cm]{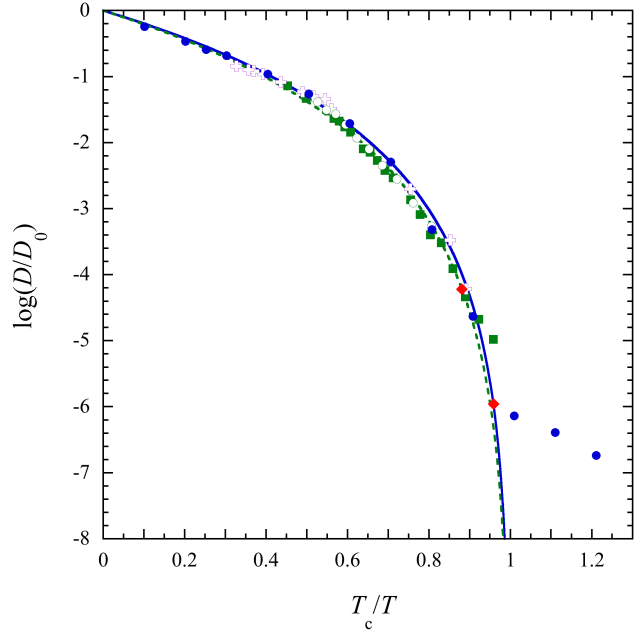}
\end{center}
\caption{(Color online) A log plot of the scaled collective diffusion coefficient $D/D_0$ versus scaled inverse temperature $T_c/T$ for a fragile liquid and a strong liquid. The symbols ($\Box$) indicate the simulation results for NV, ($\odot$) for BKS, ($\bullet$) for SW, and ($+$) for Al$_2$O$_3$. The solid line indicates the singular function given by Eq. (\ref{sltd}) for (F) and the dashed line for (S). The symbols ($\Diamond$) indicate the scaled inverse glass transition temperature $T_c/T_g$=0.958 (F) and 0.881 (S).}
\label{da}
\end{figure}
\begin{figure}
\begin{center}
\includegraphics[width=8.5cm]{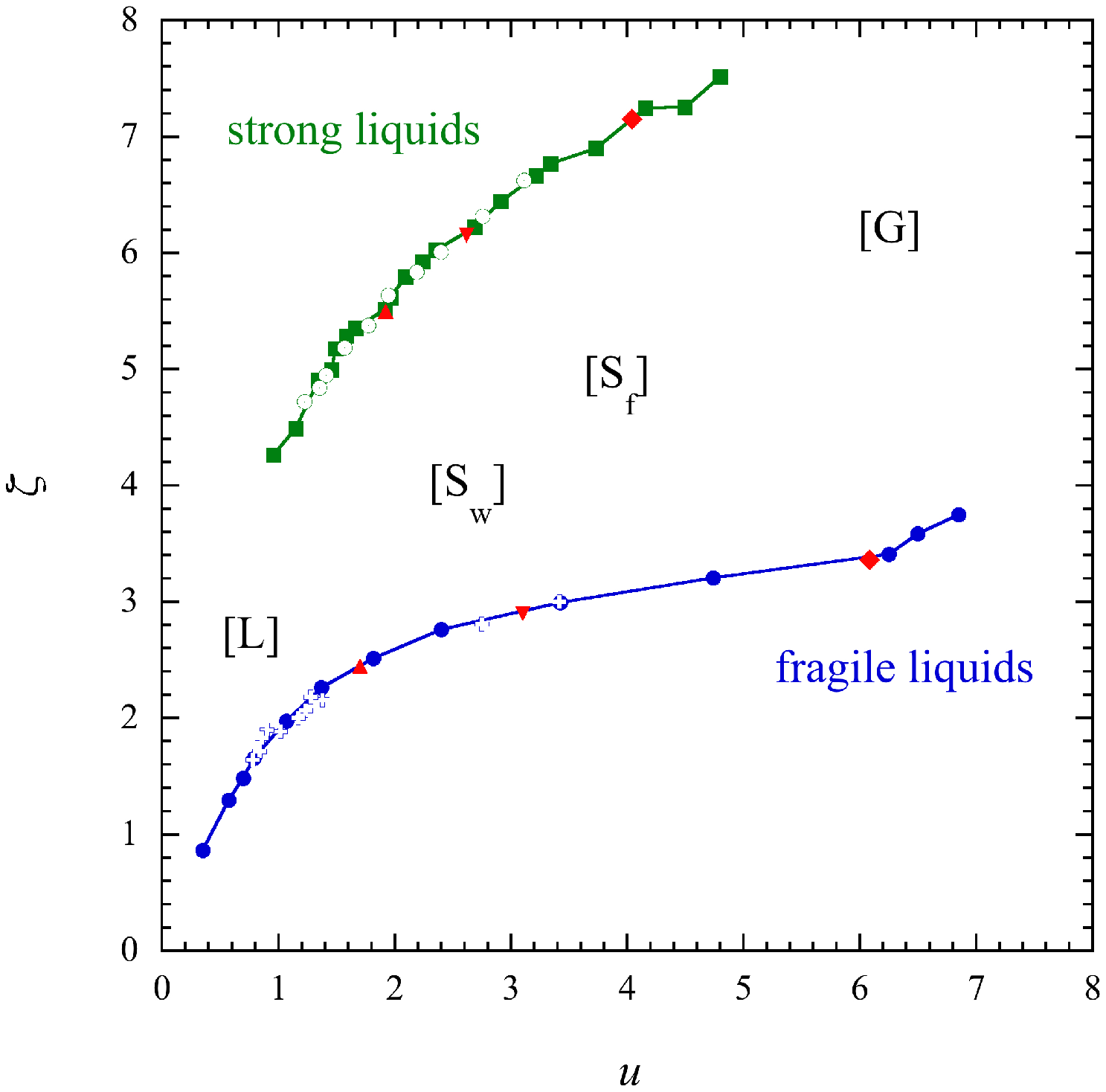}
\end{center}
\caption{(Color online) Universal function of $u$. (a) A plot of the scaled friction coefficient $\zeta$ versus $u$. The symbols ($\Box$) indicate the numerical values for NV, $(\odot$) for BKS, ($\bullet$) for SW, and ($+$) for Al$_2$O$_3$. The symbols ($\Diamond$) indicate the glass transition point $u_g$, ($\triangle$) the supercooled point $u_s$, and ($\triangledown$) the deeply supercooled point $u_f$, where the numerical values of $u_i$ are listed in Table \ref{table-2}. The solid lines are guides for eyes. The label [L] stands for a liquid state, [S$_w$] for a weakly supercooled state, [S$_f$] for a deeply supercooled state, and [G] for a glass state.}
\label{uni}
\end{figure}
Hence one can also fix the value of $\kappa$ at each value of $D$. Finally, we fix the value of $\mu$ so that the numerical solution of Eq. (\ref{kasc}) coincides with the simulation results at the $\alpha$-relaxation stage for $\tau\geq\tau_x$, where $\tau_x$ is a crossover time from the von Schweidler decay to the Kohlrausch-Williams-Watts decay \cite{toku191}. Thus, the coupled equations (\ref{kasc}) and (\ref{kappa}) can be solved self-consistently at a given value of $D$. Since all parameters in Eq. (\ref{kasc}) are uniquely determined by $D$, therefore, it turns out that there exists only one solution of Eq. (\ref{kasc}) in different liquids of a same type at a given value of $D$.
Hence it is convenient to introduce a universal parameter $u$ by \cite{toku20}
\begin{equation}
u=-\log(D(T)). \label{up}
\end{equation}

\begin{figure}
\begin{center}
\includegraphics[width=8.5cm]{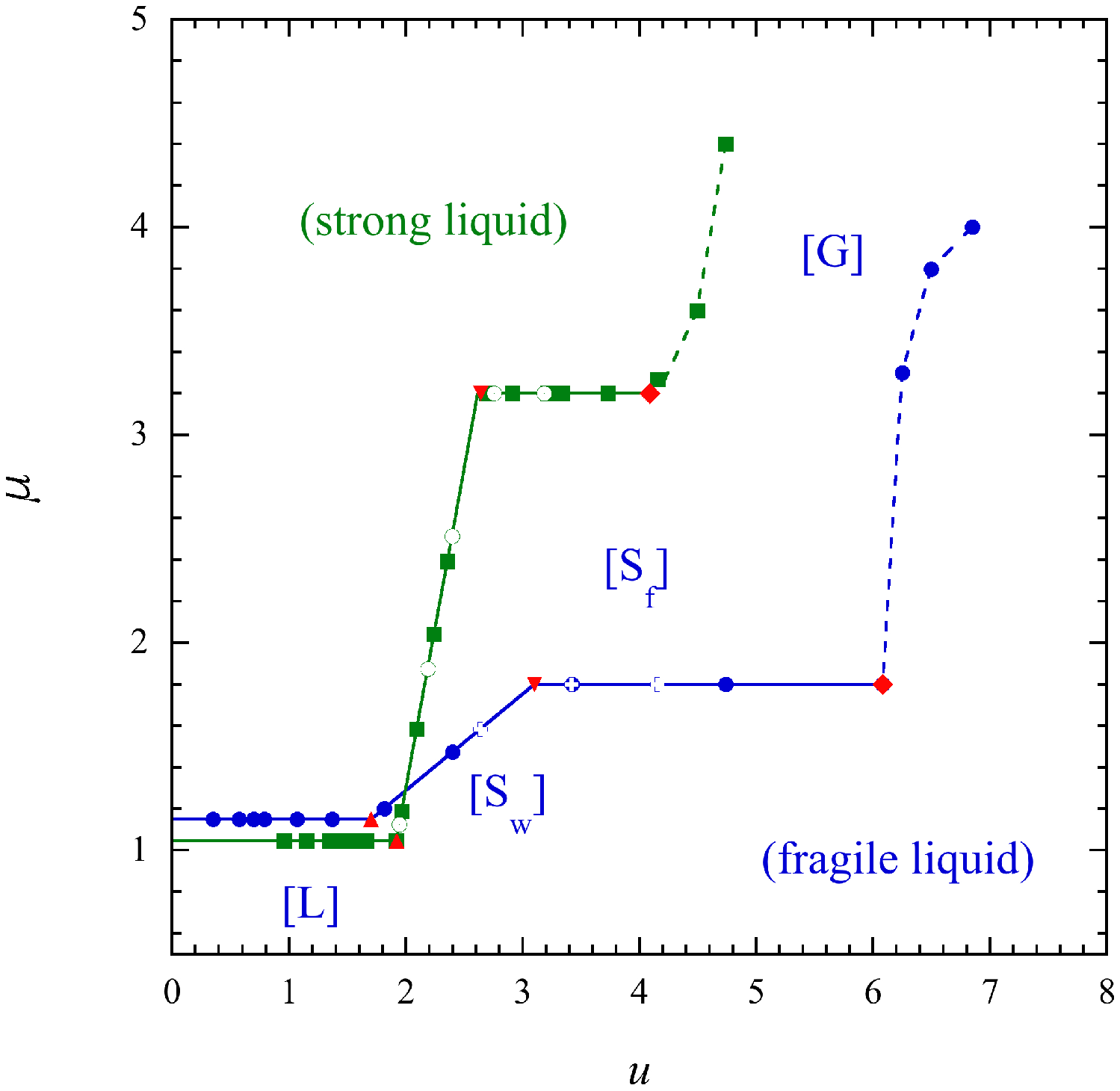}
\end{center}
\caption{(Color online) A plot of the nonlinear parameter $\mu$ versus $u$ for various liquids. The symbols ($\bullet$) indicate the fitting values for SW, ($+$) for Al$_2$O$_3$, ($\Box$) for NV, and ($\odot$) for BKS. The symbols ($\Diamond$) indicate the glass transition point $u_g$, ($\triangle$) the supercooled point $u_s$, and ($\triangledown$) the deeply supercooled point $u_f$, where the numerical values of $u_i$ are listed in Table \ref{table-2}. The solid lines are given by Eqs. (\ref{fragile}) and (\ref{strong}). The dashed lines are guides for eyes.  The details are the same as in Fig. \ref{uni}. }
\label{pfal}
\end{figure}
\begin{figure}
\begin{center}
\includegraphics[width=8.5cm]{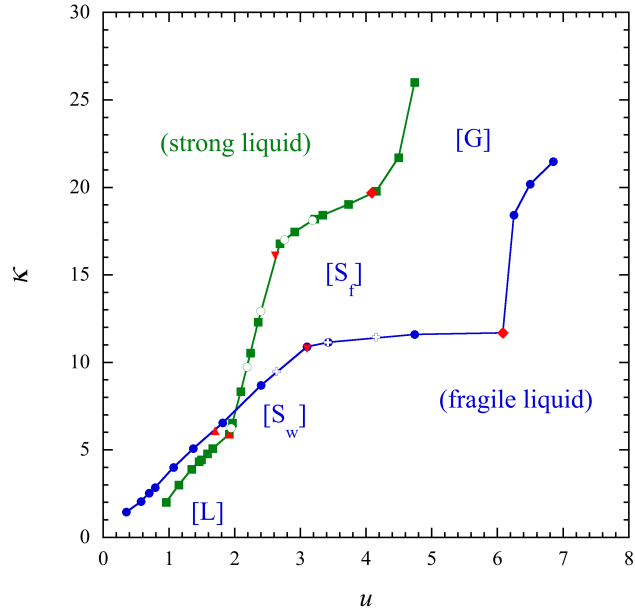}
\end{center}
\caption{(Color online) A plot of the coupling parameter $\kappa$ versus $u$ for various liquids. The details are the same as in Fig. \ref{pfal}. The solid lines are guides for eyes.}
\label{ka}
\end{figure}
\begin{table}[b]
\caption{Glass transition temperature $T_g$, supercooled temperature $T_s$, deeply supercooled temperature $T_f$, and corresponding universal parameters $u_g=u(T=T_g)$, $u_s=u(T=T_s)$, and $u_f=u(T=T_f)$ for different systems.}
\begin{center}
\begin{tabular}{cccccccc}
\hline
type & system  & $T_c/T_g$ &$T_c/T_f$ &$T_c/T_s$&$u_g$&$u_f$&$u_s$\\
\hline
fragile &SW & 0.958 &0.784 &0.567&6.082&3.102&1.705\\

\hline
strong &NV   &0.881 &0.754 &0.650 &4.040&2.620&1.920\\
\hline
\end{tabular}
\end{center}
\label{table-2}
\end{table}
As shown in Ref. \cite{toku20}, in order to find $D$ and $\zeta$ numerically, we have first taken the simulation results \cite{toku16,toku141,toku131} performed on the binary mixtures A$_{80}$B$_{20}$ with the Stillinger-Weber potential (SW) \cite{sw} as a simple example of (F) and SiO$_2$ with the Nakano-Vashishta potential (NV) \cite{nv} as a simple example of (S), where $\zeta_0(T)\simeq$ 12 (SW) and 96 (NV), and $q_m\sigma=$7.25 (SW) and 1.55 (NV). 
The numerical values of $D$ and $\zeta$ are then found from those simulation results. In Fig. \ref{da}, the simulation results for $D$ are plotted versus scaled inverse temperature $T_c/T$ together with the singular function given by
\begin{equation}
D(T)=D_0(1-T_c/T)^{\gamma}, \label{sltd}
\end{equation}
where $\gamma$ is a power exponent to be determined, $T_c$ a critical temperature, and $D_0$ a prefactor to be determined. Here the values of $D$ obtained from the simulation results \cite{toku17} performed on Al$_2$O$_3$ with the Born-Meyer potential \cite{bm} and SiO$_2$ with the Beest-Kramer-Santen (BKS) potential \cite{bks} are also shown in Fig. \ref{da}, where $\zeta_0\simeq$ 82 and $q_m\sigma$=4.25 for Al$_2$O$_3$ and $\zeta_0\simeq$ 113 and $q_m\sigma$=1.65 for BKS. From the simulation results, we thus find that $\gamma\simeq 4.317$ for (F), where $D_0\simeq 0.775$ for SW and $D_0\simeq$0.692 for Al$_2$O$_3$, and $\gamma\simeq 4.563$ for (S), where $D_0\simeq 1.500$ for NV and $D_0\simeq $1.423 for BKS. Here we should note that the power exponent $\gamma$ of (S) is slightly larger than that of (F). This situation is the same as that in $D_s$ \cite{toku13}. In Fig. \ref{uni}, the scaled friction coefficient $\zeta$ is also plotted versus $u$. In each type of liquids $\zeta$ is thus shown to coincide with each other within error. Hence the dynamics in different liquids of a same type is easily expected to coincide with each other at a given value of $u$ \cite{toku20}. 

The $u$ dependence of the nonlinear parameter $\mu$ is first obtained for SW and NV. Then, such a dependence is also checked for Al$_2$O$_3$ and BKS and is confirmed to hold also for them. In Fig. \ref{pfal}, $\mu$ is plotted versus $u$ for different liquids. Depending on a value of $u$, there exist three states, a liquid state [L], a supercooled state [S], and a glass state [G]. As shown in the previous paper \cite {toku20}, the supercooled state is further separated into two substates, a weakly supercooled state [S$_w$] and a deeply supercooled state [S$_f$]. In each state the value of $\mu$ is given for (F) and (S) by
\begin{equation}
\mu\simeq\begin{cases} 1.150, & \text{[L]\; for $u\leq u_s$}\\
 0.466u+0.355, & \text{[S$_w$] for $u_s\leq u<u_f$}\\
 1.8, & \text{[S$_f$] for $u_f\leq u\leq u_g$} \label{fragile}
\end{cases}
\end{equation}
\begin{equation}
\mu\simeq\begin{cases} 1.046 & \text{[L]\; for $u\leq u_s$}\\
3.077u-4.862, & \text{[S$_w$] for $u_s\leq u<u_f$}\\
3.2, & \text{[S$_f$] for $u_f\leq u\leq u_g$} \label{strong}
\end{cases}
\end{equation}
respectively. In [L] $\mu$ is constant since the magnitude of the equilibrium density fluctuations around an equilibrium density is small. In [S$_w$] it grows as $u$ increases since the magnitude of fluctuations becomes larger, while in [S$_f$] it is constant since the magnitude seems to be constant. Hence the system must be metastable in [S$_w$], while it must be stable in [S$_f$]. As mentioned in Ref. \cite{toku20}, those behavior in [S] must be explained by the so-called spatial heterogeneity \cite{ag65,toku97,toku99,si99,ed00,w00,rr02,ph08,voig16}. On the other hand, in [G] for $u_g<u$ $\mu$ grows rapidly as $u$ increases. This is because the system is out of equilibrium although the magnitude of the nonequilibrium density fluctuations around the averaged nonequilibrium density is small. In Fig. \ref{ka}, the coupling parameter $\kappa$ is also shown to grow monotonically as $u$ increases, except in [S$_f$] where it grows slowly. Thus, it is turned out that the strong liquids are qualitatively similar to the fragile liquids but is quantitatively different from them. Hence the dynamics of strong liquids never coincides with that of fragile liquids even at a given value of $u$.
\begin{figure}
\begin{center}
\includegraphics[width=8.5cm]{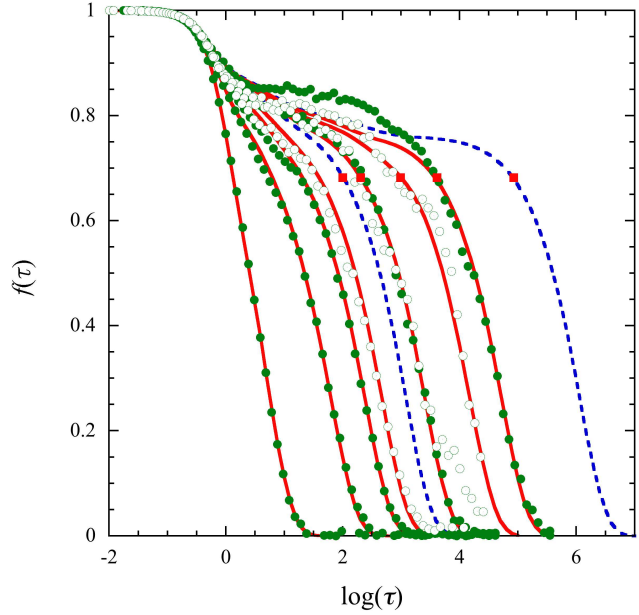}
\end{center}
\caption{(Color online) A plot of $f(\tau)$ versus $\log(\tau)$ for different values of $u$. The solid lines indicate the numerical solutions at $u=$[L] 0.700, [S$_w$] 1.821, 2.4202, [S$_f$] 3.426, 4.154, and 4.741, and the dashed lines at $u_f=3.102$ and $u_g=6.082$ (from left to right). The symbols ($\bullet$) indicate the simulation results for SW at $T$=2.00, 0.833, 0.714, 0.625, and 0.566, and ($\odot$) for Al$_2$O$_3$ at $T$=2600, 2300, and 2200(K) (from left to right). The symbols ($\Box$) indicate the crossover time $\tau_x$, where $f(\tau_x)\simeq 0.682$.}
\label{fra}
\end{figure}
\begin{figure}
\begin{center}
\includegraphics[width=8.5cm]{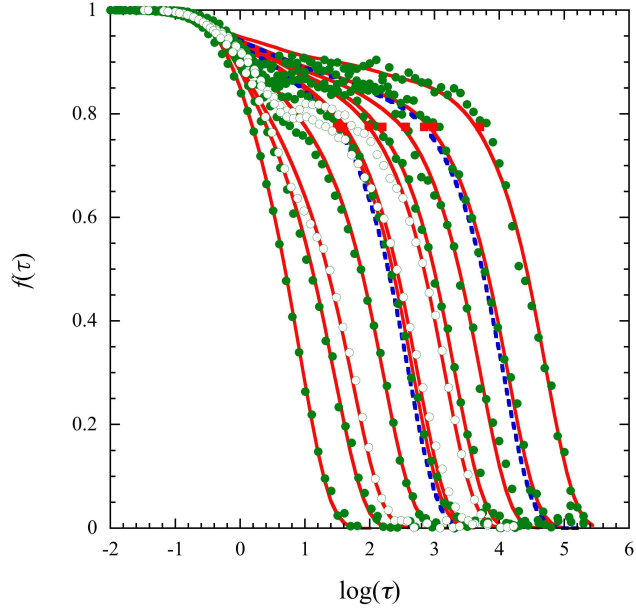}
\end{center}
\caption{(Color online) A plot of $f(\tau)$ versus $\log(\tau)$ for different values of $u$. The solid lines indicate the numerical solutions at $u=$[L] 0.962, 1.495, 1.774, [S$_w$] 2.243, [S$_f$] 2.688, 2.757, 3.184, 3.340, 3.733, [G] 4.159, and 4.802 and the dashed lines at $u_f=2.620$ and $u_g=4.040$ (from left to right). The symbols ($\bullet$) indicate the simulation results for NV at $T$=5500, 4300, 3600, 3300, 3000, 2900, 2800, and 2600(K), and ($\odot$) for BKS at $T$=4400, 3600, and 3400(K)  (from left to right). The symbols ($\Box$) indicate the crossover time $\tau_x$, where $f(\tau_x)\simeq 0.775$.}
\label{str}
\end{figure}

In Figs. \ref{fra} and \ref{str}, the numerical solutions at different values of $u$ are compared with the simulation results for (F) and (S), respectively. Since $\mu$ is given by Eqs. (\ref{fragile}) and (\ref{strong}), one can obtain the numerical solutions at any temperatures in [L] and [S], even though there is no simulation result performed at such temperatures. For example, the solutions at $T_f$ and $T_g$ are also plotted in both figures. For smaller values of $u$ in [L], the solutions are shown to coincide with the simulation results well within error. For larger values, the solutions are shown to agree with them well within error, except at $\beta$ stage for $1\leq\tau\leq\tau_x$. The deviation at $\beta$ stage is clearly seen in a supercooled state [S] and a glass state [G], while they are small in a liquid state [L]. As pointed out before, such a deviation in [S] just results from the ideal TMCT equation. Hence we should mention here that such a technical deviation must disappear if one can solve the original TMCT equation with the simplified nonlinear memory function given by Eq. (\ref{almod}) numerically. Finally, we note that the potential for SW and NV is a short-range interaction between particles, while that for Al$_2$O$_3$ and BKS is a long-range interaction. Hence errors found in the simulation results with the long-range interactions are turned out to be larger than those with the short-range interactions, especially for a longer time and also for lower temperatures. This must be just caused by a lack of an enough simulation time for the system with a long-range interaction. In order to confirm the universal properties and also to obtain precise values of universal quantities, therefore, one needs to perform more extensive molecular-dynamics simulations on various glass-forming materials at the same value of $u$ consistently.

\section{Summary}
In this paper, we have numerically solved the universal TMCT equation for $K(\tau)$ given by Eq. (\ref{kasc}) with the simplified nonlinear memory function $M(\tau)$ given by Eq. (\ref{almod}) based on the four different simulation results performed on two types of glass-forming liquids. Then, we have shown that there exists only one solution for different liquids of a same type at a given value of $u$. We have compared the solutions with the simulation results at a given value of $u$. Thus, we have shown that the numerical solutions can describe the simulation results well within error, except at $\beta$ stage because of an ideal TMCT equation. As shown in Ref. \cite{toku20}, we have also confirmed that the supercooled state is clearly separated into two substates, a weakly supercooled state [S$_w$] and a deeply supercooled state [S$_f$], where $\mu$ obeys Eqs. (\ref{fragile}) and (\ref{strong}). The $u$ dependence of $\mu$ and $\kappa$ for (F) has been shown to be qualitatively similar to that for (S) but to be quantitatively different from that. Hence the dynamics of fragile liquids has been turned out to disagree with that of strong liquids even at the same value of $u$. We mention here that such a difference is originally based on the fact that the static structure factor of strong liquids is structurally quite different from that of fragile liquids because the former has a network structure \cite{kob99,z83,m95}. Finally, we emphasize that the present universality must be useful to predict a first peak position of the static structure factor for a whole system from the experimental data. This will be discussed elsewhere.

Acknowledgments

This work was partially supported by Institute of Multidisciplinary Research for Advanced Materials, Tohoku University, Japan.

\end{document}